\documentclass[12pt]{nature}
\usepackage{amsmath,amsfonts,amssymb,amsthm,epsfig,array}
\usepackage{dsfont}
\usepackage{microtype} 
\usepackage[final]{pdfpages}
\usepackage{braket}

\usepackage[hidelinks,colorlinks,linkcolor=blue,citecolor=blue,urlcolor=blue]{hyperref}
\usepackage[titletoc,title]{appendix}
\usepackage{times}
\makeatletter
\let\saved@includegraphics\includegraphics
\AtBeginDocument{\let\includegraphics\saved@includegraphics}
\renewenvironment*{figure}{\@float{figure}}{\end@float}
\makeatother

\title{Chirality driven topological electronic structure of DNA-like materials}
\author{Yizhou Liu$^\ast$, Jiewen Xiao$^\ast$, Jahyun Koo,  Binghai Yan$^\dagger	$ \\
\normalsize{Department of Condensed Matter Physics,Weizmann Institute of Science, Rehovot 76100, Israel}
}

\begin{document} 

\maketitle 

\begin{abstract}

Topological aspects of the geometry of DNA and similar chiral molecules have received a lot of attention, while the topology of their electronic structure is less explored. Previous experiments have revealed that DNA can efficiently filter spin-polarized electrons between metal contacts, a process called chiral-induced spin-selectivity (CISS). However, the underlying correlation between chiral structure and electronic spin remains elusive. In this work, we reveal an orbital texture in the band structure, a topological characteristic induced by the chirality. We find that this orbital texture enables the chiral molecule to polarize the quantum orbital. This orbital polarization effect (OPE) induces spin polarization assisted by the spin-orbit interaction from a metal contact and leads to magnetorestistance and chiral separation. The orbital angular momentum of photoelectrons also plays an essential role in related photoemission experiments. Beyond CISS, we predict that OPE can induce spin-selective phenomena even in achiral but inversion-breaking materials.
\end{abstract}

 In chemistry and biochemistry, chirality is the geometric asymmetry of a large group of molecules with a non-superposable mirror image, either left- or right-handed. Chirality plays a prominent role in chemistry  and biology~\cite{Siegel1998} for example in enantioselective catalysis 
 and drug design. 
 In physics, chirality usually refers to the locking of spin and motion, for example in Weyl fermions\cite{Weyl1929}.  
 Although chirality represents seemingly unrelated characters in different fields, experiments~\cite{Gohler2011} have revealed an unexpected correlation between chiral geometry and electron spin. When transmitted through DNA and similar chiral molecules, electrons becomes highly spin-polarized, and their polarization depends on the chirality of the molecule. This effect is called chiral-induced spin selectivity (CISS)~\cite{Naaman2012,Naaman2019},
where a large spin polarization is induced and manipulated in unexpected ways.

Practically, the application potential of the CISS effect for spintronic devices
,chiral electrocatalysis
and  enantiomer selectivity has been demonstrated.~\cite{Naaman2012,Naaman2019}
For instance, chiral molecules were found to adsorb on a ferromagnetic substrate with different speeds that depend on both the chirality and the substrate magnetization, leading to efficient separation of enantiomers~\cite{BanerjeeGhosh2018}. When contacting to magnetic leads, the chiral molecule exhibits magnetization-dependent resistivity, i.e. magnetoresistance (MR)\cite{Xie2011,Kettner2015}. 
However, the physical origin of CISS, that is the relationship between the chiral structure and the electron spin, is still debated. 

The first characteristic feature of CISS experiments is their robustness at room temperature. The second feature is that they are dynamical phenomena that usually involve electron tunneling or electron transfer in a non-equilibrium process. Chiral molecules like DNA exhibit zero spin polarization in the ground state, and the CISS effect vanishes in equilibrium. For example, chiral separation disappears after the substrate and molecules reach thermodynamic equilibrium after a sufficient amount of time.~\cite{BanerjeeGhosh2018}
Another feature that usually attracts little attention is related transport experiments commonly have electrodes that are heavy metals (e.g., Au) or materials with strong spin-orbit coupling (SOC). In contrast, photoemission experiments on CISS use both heavy- (e.g., Au) \cite{Gohler2011,Ray1999}
 and light-element (e.g., Al and Cu) \cite{Mishra2013,Kettner2018}
substrates.

Many theories have been developed to understand the CISS effect
but no consensus has been reached so far~\cite{Naaman2019}. The developed models aim to generate the \textit{spin}-polarization via the chiral molecule to interpret both the transport and photoemission experiments.
Most models consider the chiral molecule as a spin filter and require an effective spin-orbit coupling (SOC) in the molecule to couple the electron motion and spin. However, it is known that the experimentally measured SOC is no larger than a few meV in related organic systems.
Thus it is challenging to rationalize the robustness of CISS at room temperature (26 meV), even though several scenarios were proposed to enhance the spin polarization\cite{Guo2012,Gutierrez2012,Matityahu2016,Michaeli2019,Dalum2019,Yang2019,Geyer2020,Zoellner2020}. Alternatively, Ref. \citeonline{Gersten2013} proposed that the substrate provides strong SOC and the chiral molecule filters the orbital angular momentum (OAM), to induce the spin polarization.

\begin{figure}
    \centering
    \includegraphics[width=0.6\textwidth]{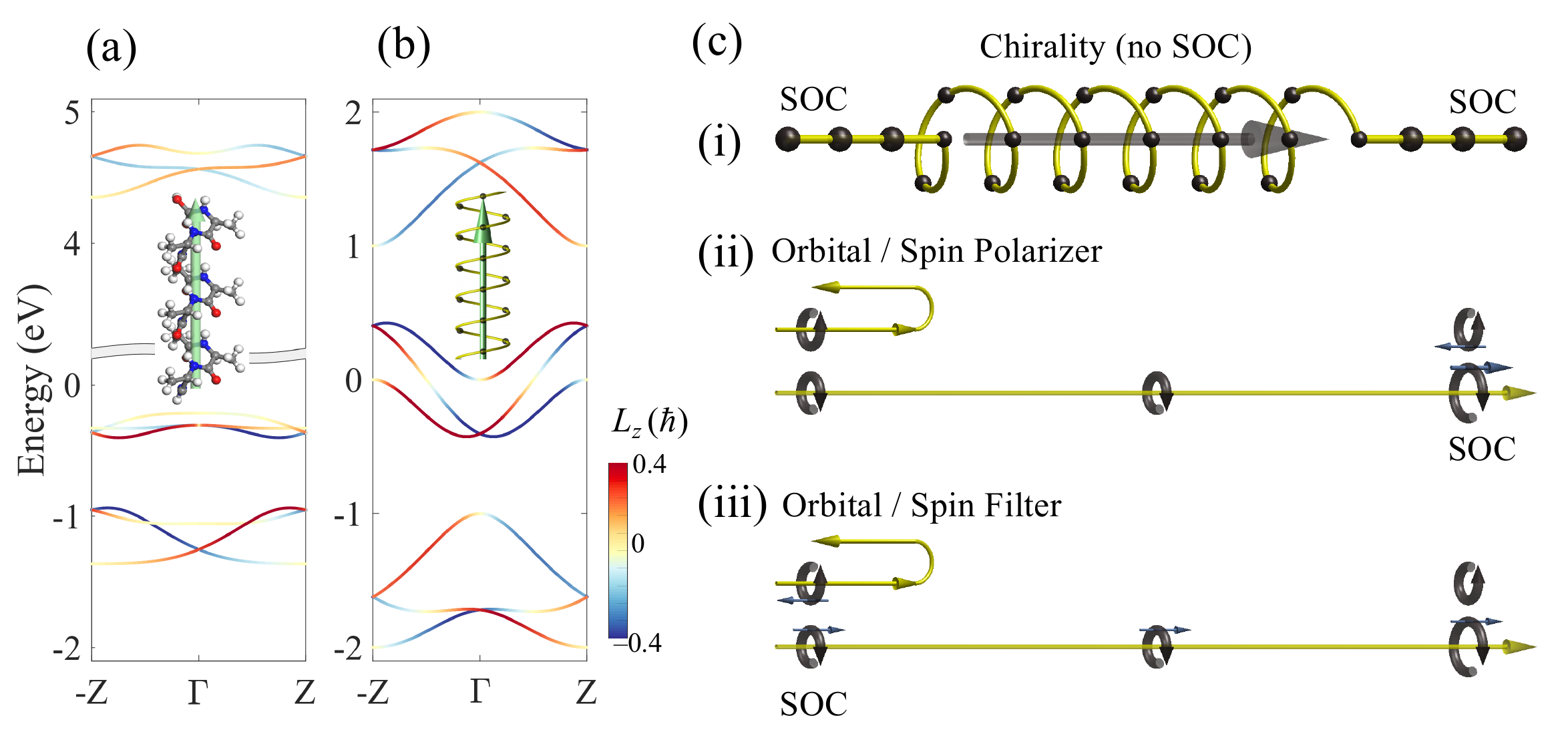}
    \caption{\textbf{The orbital polarization and the orbital texture.} 
    (a) The \textit{ab initio} band structure of the right-handed peptide $3_{10}$ helix with orbital texture. The orbital texture refers to the parallel or anti-parallel relation between orbital polarization $L_z$ and the momentum. The inset shows the atomic structure the $3_{10}$ helix where gray, blue, red, and white spheres represent C, N, O, and H atoms.  (b) The band structure of a tight-binding model of the helix. The helix has a three-fold screw rotation (see inset), same as the $3_{10}$ helix. (c) Illustration of the orbital polarization effect in transport. (i) The helix (small black spheres) connects two leads that are linear atomic chain (large black spheres). The spin-orbit coupling (SOC) exists only in leads but not in the chiral helix. At given energy, the orbital gets polarized to $+L_z$ as the electron transmits through the chiral molecule. 
    The oribtal texture leads to the orbital polarization effect (OPE), in which the chiral molecule plays roles as both an orbital filter and an orbital polarizer. (ii) The orbital polarizer. When electrons run into the right lead where SOC exists, the orbital polarization induces spin polarization. The half circles with arrows represent the $\pm L_z$ orbital. Thin arrows represent the spin.  The larger orbital and spin stand for the orbital and spin polarization, respectively, are in the right lead. (iii)The orbital filter.  The chiral molecule filters the $+L_z$ state but suppresses the $-L_z$ state injected from the left lead. The yellow curves illustrate the scattering trajectory with arrows. If the spin is pre-locked to the orbital in the left lead, transmitted electrons become spin-polarized due to the orbital filtering. We note that the multiple-mode leads allow the emergence of spin polarization in the two-terminal conductance despite the presence of the time-reversal symmetry.}
    \label{figure1}
\end{figure}

The chiral electronic structure of Weyl semimetals\cite{Yan2017,Armitage2017} inspires us to explore the band structure topology of DNA and similar chiral molecules. In this work, we reveal an ubiquitous topological orbital texture in the chiral lattice and propose a mechanism that the chiral molecule acts as an orbital polarizer and an orbital filter, rather than a spin filter, in the CISS effect. The orbital polarization effect (OPE) does not require SOC from the molecule and is robust against temperature fluctuations. The orbital refers to the orbital angular momentum (OAM) of the wave function. By calculations with the Landauer-Büttiker formalism, we demonstrate that electrons become orbital-polarized after transmission through the chiral molecule, and the polarization depends on the chirality. 
The OPE means that the chiral molecule act as both an orbital polarizer and an orbital filter, as illustrated in \textbf{Figure 1c}. The former is usually stronger than the latter.
The orbital polarizer occurs when electrons transmit from the chiral molecule into the lead.  The orbital-polarization induces spin polarization in the lead because of SOC .
The orbital filter happens when electrons propagate from the lead into the molecule. The chirality filters the orbital and subsequently selects the spin {(similar to the mechanism in Ref.\citeonline{Gersten2013})}, because orbital and spin are pre-locked by SOC in lead. 
Furthermore, when the lead is magnetic (spin-polarized), it is also orbital-polarized because of SOC. The electron tunneling is affected by whether the orbital polarization of the lead matches that of the molecule, leading to chirality and magnetization dependent MR. This also induces different adsorbance / desorbance speeds, which depend on the charge transfer efficiency of molecules with opposite chirality on the magnetic substrate, resulting in chiral separation. We rationalize CISS-induced transport phenomena (e.g., MR and the magnetic chiral selection) without requiring the presence of SOC in the molecule. We point out that the heavy metal lead plays the role of a spin-orbit translator that converts the orbital polarization into the spin polarization.
{For photoemission experiments, we propose that the measured magnetization corresponds to the global OAM besides the spin of photoelectrons, to rationalize the insensitivity to the substrate SOC.}
Furthermore, we predict that the OPE can lead to spin-selective effects in achiral but non-centronsymmetric molecules and solids.


\subsection{Topological Orbital Texture}

The band structure of a chiral lattice exhibits a topological feature that we call the orbital texture. We take a periodic chain of the right-handed peptide $3_{10}$ helix as an example. It is a typical secondary structure found in proteins and polypeptides, which has also been studied in a CISS experiment \cite{Tassinari2018}. The helix exhibits a three-fold screw rotation around the $z$-axis. This symmetry induces Dirac-like band crossings at the zone center ($\Gamma$) and boundary($\pm Z$), as shown in the \textit{ab initio} band structure in \textbf{Figure 1a}. It always sticks three bands together as a general consequence of a nonsymmorphic symmetry in the band structure topology~\cite{Parameswaran2013}.
Thus, the nature of a chiral band structure involves multiple bands, beyond the one-band description. Here, SOC is ignored in the band structure since it is negligibly small.

A salient feature in the band structure is the orbital polarization $L_z$. The $L_z$ refers to the atomic OAM. 
It is worth stressing that the OAM represents the self-rotation of the wave functions around atomic centers and is not a conserved quantity. 
To respect time-reversal symmetry (TRS), $L_z$ exhibits opposite signs at the $k_z$ and $-k_z$ points, as shown in \textbf{Figure 1a}. Similar to the chirality of Weyl fermions,
$L_z$ and $k_z$ are always parallel or anti-parallel, which depends on the molecule chirality. Different from the Weyl point that exhibits infinitely large polarization, the $L_z$ vanishes to zero at $\Gamma$ and $\pm Z$ because of TRS. According to the symmetry analysis, we point out that such an orbital texture is ubiquitous for a chiral lattice with and without a helix structure. Even when the screw rotational symmetry vanishes in a non-helical structure, the orbital texture still preserves because of the IS-breaking. 
Because of the orbital texture, oppositely propagating electrons carry opposite orbital polarization, resulting in the OPE discussed in the following.

\subsection{Orbital Polarization Effect in Transport}

Previous theoretical studies \cite{Michaeli2019,Guo2012,Gutierrez2012,Matityahu2016,Yang2019,Zoellner2020}
usually map the dynamic CISS process to a transmission problem between two achiral leads through a chiral molecule that exhibits effective SOC. The spin polarization of the transmitted electrons is evaluated as evidence of the CISS. In this work, we adopt the same transmission model but remove SOC from the chiral molecule. 
As illustrated in \textbf{Figure 2a}, we add a linear part with atomic SOC between the chiral molecule and both leads, to simulate the fact that chiral molecules commonly contact to the noble metal substrate.
Both leads and the chiral molecule have no SOC. Achiral linear chains represent leads and the SOC part. The tight-binding Hamiltonian of the whole system (both leads and the center region) is constructed with atomic $p$-orbitals ($p_x, p_y, p_z$) and the hopping parameters in the same way as the band structure calculation in \textbf{Figure 1c}.

We calculated the conductance ($G$) using the Landauer-B\"uttiker formulism~\cite{Buttiker1986}. The spin and orbital-polarized conductances of transmitted electrons are represented by $G_{S_z}$ and $G_{L_z}$, respectively (please refer to the Methods section). Corresponding spin and orbital polarization ratios are $P_{S_z}=G_{S_z} / G$ and $P_{L_z}=G_{L_z} /G$, respectively. To demonstrate the CISS, we will show that electrons go through the chiral molecule and get spin-polarized, which is caused by OPE. The existence of the two-terminal MR, i.e. $G_{L \rightarrow R} (+M) \neq G_{L \rightarrow R}(-M)$, requires the dephasing to leak electrons into virtual leads\cite{Buttiker1986b}. Here $G_{L \rightarrow R} (+M)$ represents the conductance from the left to the right leads, where the leads' magnetization is $+M$.

The orbital texture induces the OPE as electrons move through the chiral molecule. 
The chiral molecule acts as both an orbital filter and an orbital polarizer. If we approximate $L_z$ as conserved, then $L_z$ does not flip when traveling through the molecule. We can regard the chiral channel as an orbital filter that only allows a given $L_z$ to pass,  as illustrated in \textbf{Figure 1c}. The orbital filter is essential at the interface between the chiral molecule and the left lead, where the incident orbitals are filtered. However, $L_z$ is not a conserved quantity in the molecule and will flip in the transmission. The $G_{L_z}$ also includes substantial orbital-flip contribution due to the transmission from $p_{-}$ and $p_z$ states at left lead to $p_{+}$ at right (see the channel-specific conductance in SM). It is interesting to see that the chiral channel can even polarize the $\ket{L_z=0}$ state. The orbital polarizer is vital when electrons run from the molecule into the right lead. These emitting electrons can induce intense orbital polarization in the right lead. {(see more in the supplementary Figs. S7 and S8)}

The chiral molecule exhibits preferred transmission for electrons with the orbital polarization parallel or anti-parallel, which is determined by the chirality with respect to the transmitting direction at a specific energy. Our conductance calculations confirm the OPE, as indicated by the nonzero $G_{L_z}$ in \textbf{Figure 2b}. We turn off the SOC in the whole device and observe no spin polarization in the conductance since the motion and spin do not separate. We recall that the linear chain exhibits no orbital texture. The OPE is only due to the orbital texture in the chiral region. 

\begin{figure}
    \centering
    \includegraphics[width=0.8\textwidth]{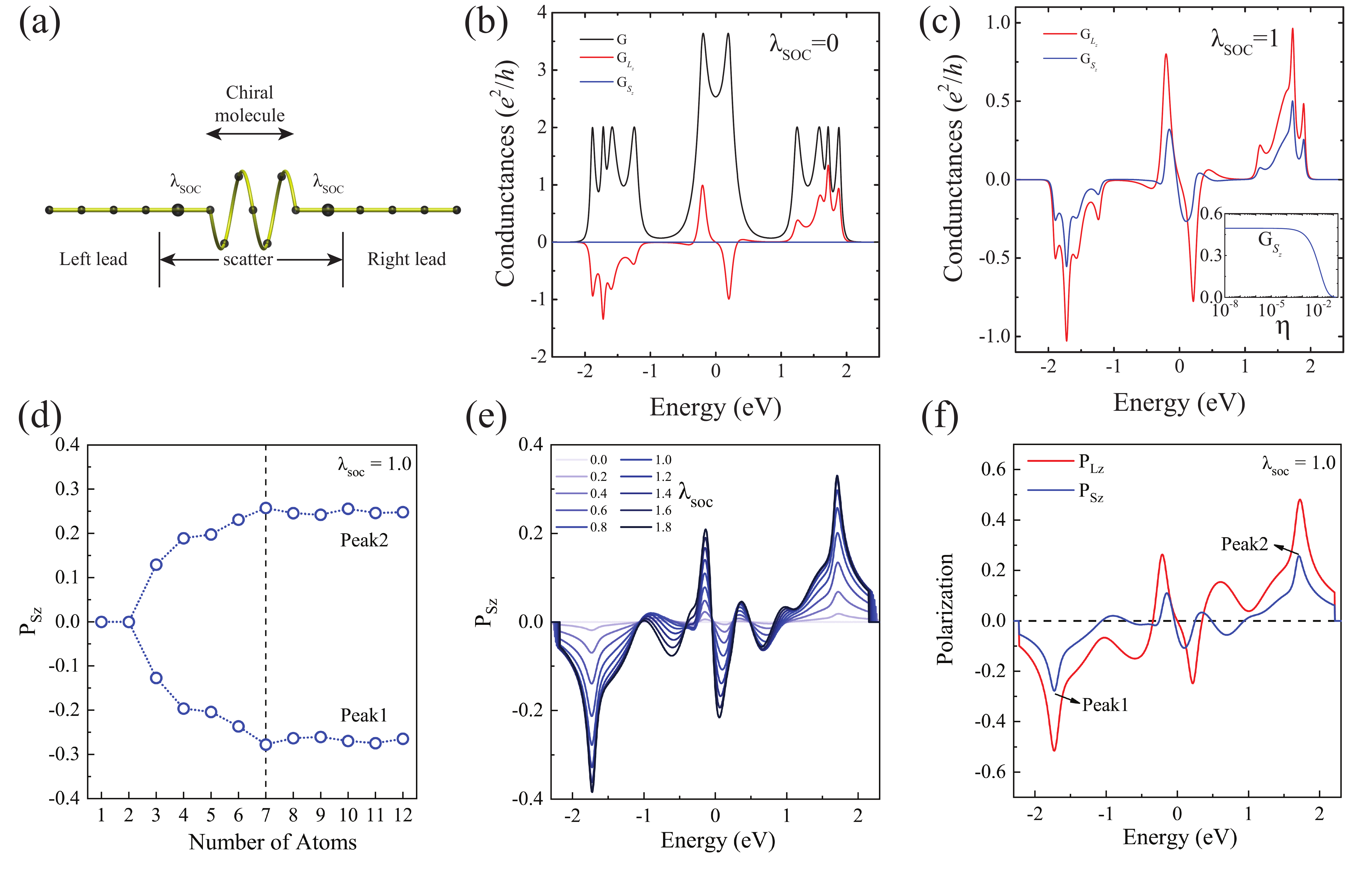}
    \caption{\textbf{Orbital polarization and spin polarization in the conductance.} (a) The transport model includes two linear leads and the chiral molecule with two helical units long. The SOC is only added to the intermediate site between the chiral molecule and the lead. (b) The orbital polarization $G_{L_z}$ exists while the spin conductance $G_{S_z} = 0$ for $\lambda_{SOC} = 0$. $G$ is the total conductance. (c)(e)\&(f) $G_{S_z}$ and the spin polarization rate $P_{S_z}$ increases as turning on SOC. The $G_{S_z}$ (peak value around 1.8 eV) dependence on the dephasing term $\eta$ is shown in the inset of (c).
    The SOC translates the orbital polarization to the spin polarization. 
    If reversing the chirality of the molecule, the orbital texture gets inverted, and thus, the spin polarization can be flipped. (d) Peaks of $P_{S_z}$ [noted in (f)] increase quickly as increasing the length of the chiral molecule and get almost saturated after the number of atoms is 7 [the same length as the model shown in (a)]. We note that $N=1,2$ are actually achiral segments and thus corresponding $P_{S_z}$ is zero. No dephasing is included in calculations except the inset of (c).}
    \label{figure2}
\end{figure}

With SOC from the contact, the OPE eventually leads to the spin polarization. We turn on the atomic SOC ($\lambda_{SOC}$) in the short linear chains. For simplicity, we attach the SOC to both leads to make them symmetric. As shown in \textbf{Figure 2}, the spin polarization ($P_{S_z}$) increases as $\lambda_{SOC}$ increases. At given $\lambda_{SOC}$, one finds that $G_{S_z}$ is roughly proportional to $G_{L_z}$. 
Increasing the atomic number ($N$) in the chiral chain, the calculated $P_{S_z}$ first increases and soon gets nearly saturated after $N=7$. In reality, the critical length depends on the material details. The region of increasing $P_{S_z}$ can be used to interpret the length-enhanced spin polarization in the experiment~\cite{Gohler2011,Kettner2015}.

The OPE is a robust effect compared to the temperature because the order of magnitude of the orbital texture is comparable to the bandwidth (e.g.  $\sim 0.5$ eV for the $3_{10}$ helix in \textbf{Figure 1a}) and thus is much larger than room temperature. Given that SOC in the heavy metal lead is also of a similar magnitude, the resultant spin polarization in CISS becomes a robust phenomenon in ambient conditions.

\subsection{Magnetoresistance and Magnetic Chiral Separation}

\begin{figure}
    \centering
    \includegraphics[width=0.6\textwidth]{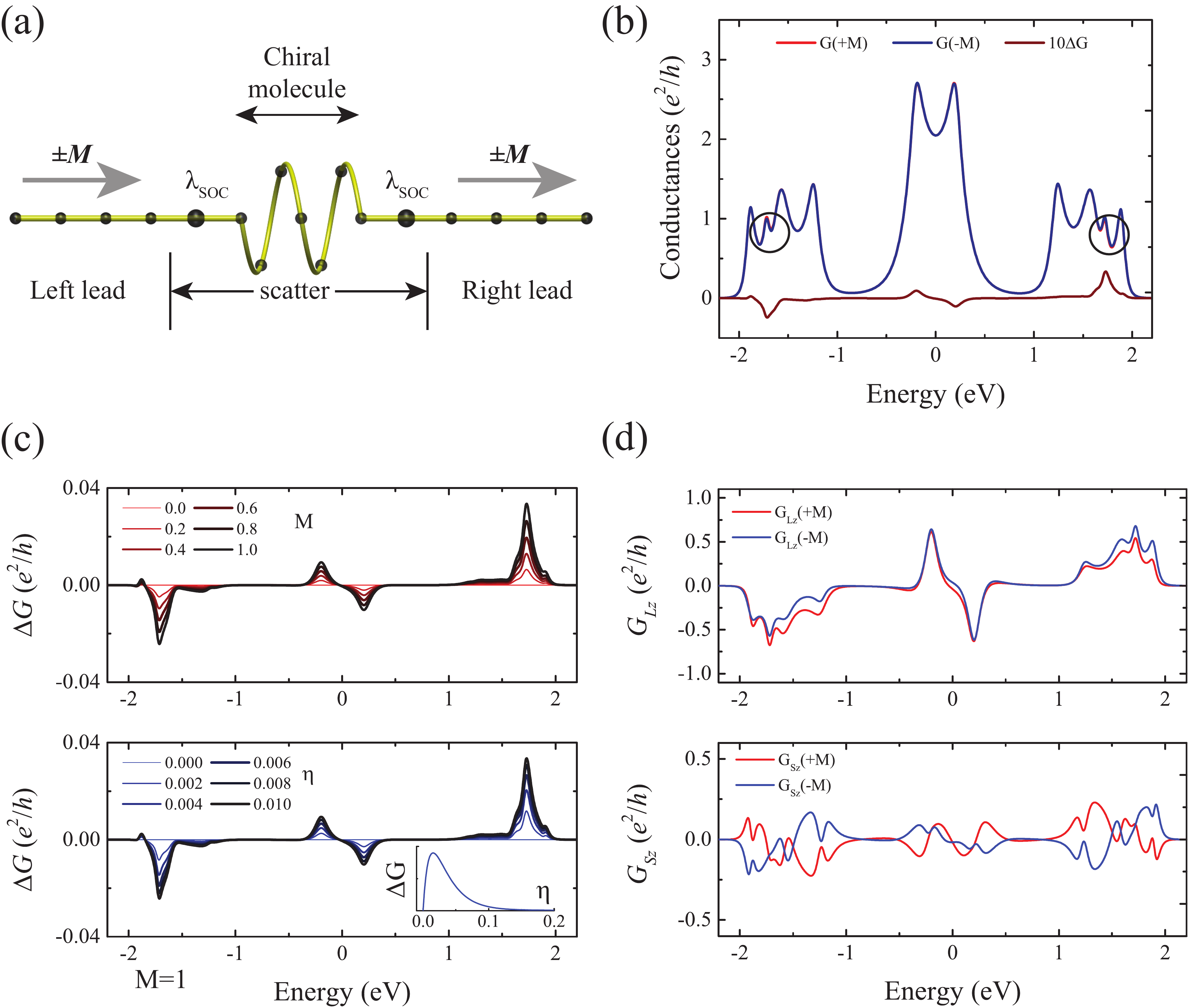}
    \caption{\textbf{Magnetoresistance with magnetic leads.} (a) The device model with two magnetized ($\pm \mathbf{M}$) leads. We employ the same two-terminal model and add an exchange field ($M$) along the $z$-direction to the spin components in both leads. The intermediate SOC regime mimics the noble metal part in experiment. 
    (b) The total conductance varies when flipping the lead magnetization. (c) The change of the conductance $\Delta G = G(-\mathbf{M})-G(+\mathbf{M})$ increases as increasing the magnitude of $\mathbf{M}$ and the dephasing term $\eta$. The inset of the lower panel shows the $\eta$-dependence of $\Delta G$ (the peak at about 1.8 eV) in a larger scale. (d) As flipping $\mathbf{M}$ from + to --, $G_{S_z}$ changes sign in the general energy window, leading to changes of $G_{L_z}$. The increase (decrease) of $G_{L_z}$ induces the enhancing (suppressing) of $G$. Here, we include a finite dephasing parameter. 
    In the coherent two-terminal measurement ($\eta =0$), the reciprocity theorem requires $G_{L \rightarrow R}(M) = G_{L \rightarrow R}(-M)$ ($\Delta G = 0$). However, the virtual lead ($i\eta$) releases this constrain, resulting in nonreciprocal MR.  If the dephasing is too large, electrons are completely incoherent and feel no OPE in the transport. This explains the decreasing of MR for large dephasing. 
   We note that the global Onsager's reciprocal relation, $G_{L \rightarrow R}(M) = G_{R \rightarrow L}(-M)$ always holds (see SM). We set $\lambda_{SOC} = 0.5$ in all these calculations.}
    \label{figure3}
\end{figure}

The orbital polarization can also be used to rationalize magnetoresistance (MR) experiments
\cite{Xie2011,Kettner2015,Zwang2016,Liu2020} 
and chiral separation by the magnetic substrate~\cite{BanerjeeGhosh2018,Metzger2020}.
In MR experiments, a gold nanoparticle was included between one lead and the chiral molecule and the other lead is a magnetic element such as nickel. Switching the lead magnetization induces the change of resistance. For chirality selection, the substrate is a ferromagnetic Co film covered by a thin layer (several nanometers) of gold. Molecules with opposite chirality get adsorbed to the substrate at different speeds, leading to the separation of chiral enantiomers.  In the transient state when the molecule approaches to the metal surface, charge redistribution occurs between the molecule and the substrate\cite{Tassinari2018}. 
The speed of the effective charge transfer, which is a quantity similar to the conductance, characterizes the speed of the adsobance. 
Therefore, we expect to gain useful insights both for the adsorbance and MR from the  conductance calculations. 

When injecting spin-polarized electrons from the substrate, the gold regime becomes spin-polarized and also orbital polarized because of SOC. Then the orbital direction is locked with the magnetization direction. If this orbital matches the following OPE in the chiral molecule, the total conductance is large but is small otherwise. Consequently, different chirality and magnetization can lead to different MR and speed of adsorbance. The MR is shown in \textbf{Figure 3}. The total conductance changes when switching the sign of magnetization and exhibits a unidirectional feature (or nonreciprocal MR) at given chemical potential. Here the dephasing is necessary. The nonreciprocal MR is consistent with the fact that the differential conductance changes when reversing $M$ in experiment.

The unidirectional conductance can also rationalize the chiral separation. We note that the adsorbance and desorbance correspond to opposite charge transfer directions between the substrate and the chiral molecule. 
According to the unidirectional conductance, a chiral molecule releases slower (faster) from the surface if it adsorbs faster (slower). Both adsorbance and desorbance guarantees that a substrate with certain magnetization attracts one chirality faster than the opposite chirality. 

It has been argued \cite{BanerjeeGhosh2018,Santra2019,Ziv2019,Cuniberti2020} that the exchange interaction between the molecule and the substrate induces the selective adsorbance by considering a transient chirality-dependent spin polarization in the molecules. 
However, Refs.~\citeonline{BanerjeeGhosh2018,Santra2019,Ziv2019} have pointed out that the transient spin polarization is induced by the charge redistribution among the molecule and the substrate, which we call the charge transfer. We state that the charge transfer is already chirality-dependent as the driving force of the chiral selection. Compared to the charge transfer, the spin polarization and exchange coupling are higher-order effects if they exist.

\section*{Discussion}
\subsection{Unidirectional Conductance and Electric Magnetochiral Anisotropy}

In the CISS-induced MR, as discussed above, the lead magnetization and the chirality together pick up one direction, along which the current flow is favored in the device. The conductance ($G$) and resistance ($R$) can be described to the leading order as,
\begin{align}
    G (\mathbf{M},\mathbf{I}) &= G^{0} + G^{\chi}\mathbf{M} \cdot \mathbf{I}\label{Conductance} \\
    R (\mathbf{M},\mathbf{I}) &= R^0 - R^{\chi} \mathbf{M} \cdot \mathbf{I}
\label{Resistance}
\end{align}
where $R^0 = 1/G^0$, $R^{\chi}=G^{\chi}/(G^0)^2$, $\mathbf{M}$ stands for the magnetization in the lead, $\mathbf{I}$ for the current, $G^\chi$ for the chirality ($\chi$) determined conductance ($G^\chi = -G^{-\chi}$). 
$G^0$ is the ordinary conductance while $G^{\chi}$ characterizes the unidirectional conductance. 
For the I-V relation, we obtain 
\begin{equation}
    I = G^0 V / (1-G^\chi M V) \approx G^0 V + G^0 G^{\chi} M V^2
    \label{IV}
\end{equation}
where the sign of $M$ can be $\pm$. It indicates that the CISS-induced MR can be probed by a two-terminal measurement in the nonlinear regime. 
{Equation~\ref{IV} agrees with the nonlinear MR discussed in Refs.\citeonline{Yang2019,Yang2019b} which are based on a symmetry analysis without considering the microscopic mechanism of CISS.}
Equation~\ref{IV} agrees with the general nonlinear I-V profile and explains the MR due to flipping the lead magnetization in experiments. 
However, it shows an asymmetric feature between $\pm V$ while experiments are approximately symmetric, which deserves future investigations. 
In addition, the unidirectional conductance can lead to current rectification, which can be applied for photon-detection or energy-harvesting.

The nonreciprocal conductance described by Eqs.~\ref{Conductance} \&~\ref{Resistance} is reminiscent of the electrical magnetochiral anisotropy (EMChA) discussed in literature\cite{Rikken2001,Tokura2018}.
Corresponding resistance change ($\Delta R$) of a chiral conductor (chirality $\chi=\pm 1$) subject to a longitudinal magnetic field $\mathbf{B}$ is expressed as $\Delta R  = R_0 \chi \mathbf{B} \cdot \mathbf{I}$. This effect was heuristically derived by generalizing Onsager's reciprocal theorem into the nonlinear regime~\cite{Rikken2001}. 
Our work reveals an unambiguous link between EMChA and the CISS-induced nonreciprocal MR. 
The EMChA refers to the conductor regime where both the TRS- and IS-breaking occur. In comparison, the CISS-induced MR is induced by the IS-breaking in the conductor but the TRS-breaking in leads. By generalizing this symmetry condition in a two-terminal device, we can distribute the IS- and TRS-breaking to any of three regimes, including two leads and the conductor, to induce the nonreciprocity. We note that the magnetic leads can also be antiferromagnetic since some noncollinear antiferromagnets can also generate spin-polarized current ~\cite{Zelezny2017}. Also, the OPE provides one possible microscopic scenario for EMChA by revealing the role of the quantum orbital and SOC. 

\begin{figure}
    \centering
    \includegraphics[width=1\textwidth]{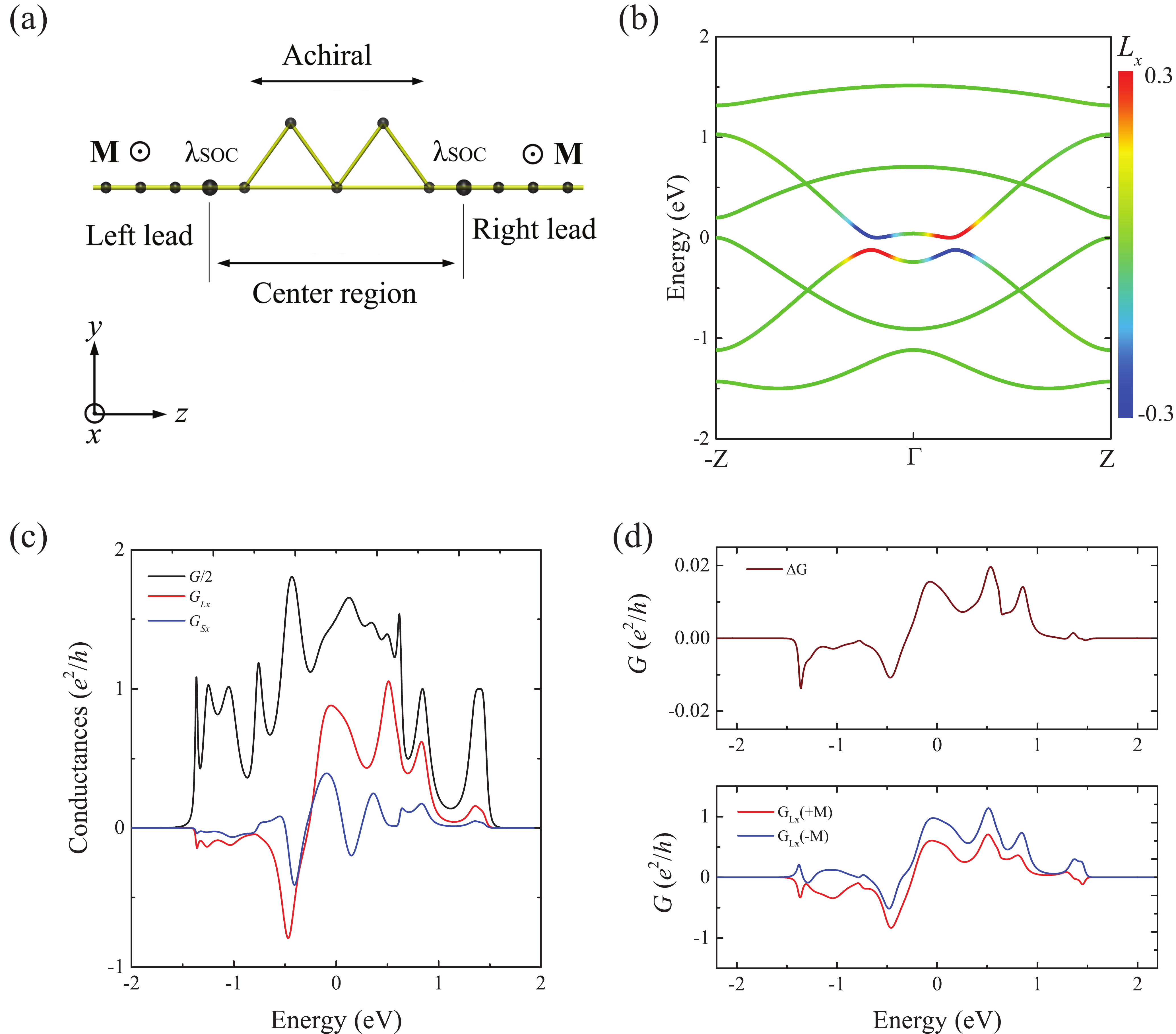}
    \caption{The orbital and spin polarization in an achiral system. (a) The device structure. The center region represents a molecule that has reflection symmetry but breaks inversion symmetry. (b) Corresponding band structure with $L_x$ orbital polarization. (c) The conductance is both orbital and spin-polarized when leads are nonmagnetic. The spin-polarization is along the $x$ direction rather than the $z$ axis.
    (d) It also exhibits magnetoresistance (the dephasing $\eta=0.005$) as switching on the magnetization from the $+x$ to $-x$ directions in leads. }
    \label{fig:my_label}
\end{figure}

\subsection{CISS in Photoemission}

{The detectable spin polarization induced by chirality requires three ingredients: SOC, TRS-breaking and chirality-induced inversion-breaking. }
As discussed above, the required SOC comes from heavy metal leads. 
The $\lambda_{SOC}$ dependence in contacts can be examined to verify our OPE theory in transport experiments. 
{However, some photoemission experiments include negligible SOC in the substrate or the molecule (except in the Mott polarizer). In this case,
the missing SOC excludes that photoelectrons are \textit{spin}-polarized }

Under intense light irradiation, the substrate ejects electrons into the vacuum through a layer of chiral molecules. In general, OAM provides the essential ingredient for CISS-induced phenomena. But OAM behaves differently in photoemission and transport systems. The OAM is quantized for photoelectrons in vacuum but not conserved for transport electrons in the lattice. The OAM can be directly measured in the  photoemission detector. However, it first needs converting into spin by SOC and then probing by MR.

We first assume the substrate has negligible SOC and then ignore the spin of electrons. After being scattered by chiral molecules, the outgoing photoelectrons (free electrons) are not a simple plane wave ($e^{ikz}$) but carry a global OAM, which can be expressed as $e^{i k z}e^{i l \phi}$, where $l$ represents the OAM and $\phi$ is the azimuthal angle. 
The global OAM represents the self-rotation of the \textit{free-electron} wave packet and is a quantized value ($l=\pm 1,\pm 2, ...$)\cite{Bliokh2007}, different from the local OAM (non-quantzied, bounded to the atomic potential) in the lattice discussed above. 
In the field of electron vortices \cite{Lloyd2017}, this OAM has been extensively studied. 
For example, experiments found that an electron plane wave obtains quantized OAM after it passes through a spiral phase plate formed by stacking graphite thin films\cite{Uchida2010}. In CISS, chiral molecules play the same role as a spiral phase plate for photoelectrons.

In a Mott polarizer, the relativistic effect generates an effective magnetic field and detects the total magnetic moment of photoelectrons. 
The total magnetic moment includes both the OAM and spin. Its sign is determined by the OAM, because the amplitude of $l$ is larger than spin 1/2.
Therefore, the polarization measured in the detector is less sensitive to the possible spin polarization induced by the substrate SOC, compared to transport experiments.
Furthermore, we propose that experiments should measure the OAM by using similar techniques employed in the vortex electron beam studies.

\subsection{Beyond the Chiral Structure}

We point out that the OPE can also generate nontrivial spin-transport phenomena in non-helical and even non-chiral systems. The OPE is caused by orbital texture, which only requires the IS-breaking if TRS exists. The chiral structure represents a strong case of the IS-breaking. Furthermore, the induced orbital polarization may differ from the current direction, depending on the way of IS-breaking.

Take an achiral chain for example, see \textbf{Figure 4a}. It is periodic along the $z$ axis and has mirror reflection for $x$ (point group $C_{2v}$). The mirror symmetry forces orbitals $L_z, L_y$ to vanish but allows for the existence of $L_x$. In the band structure, the orbital texture refers to the locking between $L_x$ and $k_z$ (see \textbf{Figure 4b}). If leads are nonmagnetic, OPE induces nonzero $G_{S_x}$ rather than $G_{S_z}$ in the presence of SOC from the contact. If leads exhibit magnetization along the $x$ direction, OPE induces the MR by reversing the magnetization. This model shows that the spin polarization does not necessarily align with the current flow. For a general noncentrosymmetric material (chiral or achiral), the orbital polarization depends on the specific symmetry. Therefore, we can engineer the geometric atomic structure to tailor the direction and magnitude of the spin-polarization. 

\section*{Outlook}

Our orbital polarization effect (OPE) brings a missing block, the orbital degree of freedom, to understand the consequence of chiral atomic structures. We rationalize transport and photoemission experiments with the local orbital angular momenetum (OAM) and global OAM, respectively.
In transport, the orbital polarization effect circumvents the weak spin-orbit coupling (SOC) in organic molecules and explains the robustness of the CISS-induced phenomena, by the intense orbital texture in the molecule and the large SOC in the lead.  The orbital texture provides an insightful quantity in the band structure to estimate the CISS effect for real materials. 
Additionally, we point out that dephasing is required to induce the magnetoresistance and magnetic chiral selection. 
The nonreciprocal conductance can lead to current rectification and may be applied for photodetection or energy harvesting. 
From the OPE, we can deduce the electrical magnetochiral anisotropy independently, which refers to the nonreciprocal magnetoresistance observed in the solid-state materials. Beyond helical molecules and even beyond the chiral structure, the OPE paves a way to manipulate the spin polarization by engineering the atomic structure in general noncentrosymmetric materials. Since chirality is a common feature of many chemical and most biochemical systems, the scope of OPE may be larger than one can imagine from the CISS, which calls for further investigations. For photoemission, we pointed out that the CISS polarization can be dominated by the global OAM of photoelectrons, insensitive to the SOC in substrate.
Our work may provide a topological perspective to understand the fundamental role of chirality in biological
and chemical systems. 

\section*{Methods}
(i) Density-functional theory. We calculate the band structure of $3_{10}$ helix and $\alpha$-helix by the density-functional theory within the generalized gradient approximation\cite{perdew1996generalized} using the the Vienna Ab initio Simulation Package (VASP) \cite{kresse1996efficient}. The orbital moment $L_z$ is extracted from the phase-dependent atomic-orbital projection of the Bloch wave function. Information for the transport calculations can be found in the SM.

(ii) Tight-binding model and orbital texture. Without loss of generality, the OAM operator $\hat{L}_z$ in the $p_{x,y,z}$ basis, is known to be,
\begin{equation*}
\hat{L}_z = 
\begin{bmatrix}
0 & -i & 0\\
i &  0 & 0 \\
0 & 0 & 0 
\end{bmatrix}
\end{equation*},
where we omit the index for atomic sites.
It has three eigen states $p_{\pm}\equiv(p_x \pm i p_y)/\sqrt{2}, p_z$ with $L_z = \pm 1, 0$, respectively. Because of the inversion symmetry(IS)-breaking, the Bloch wave function is allowed to exhibit nonzero OAM at the finite momentum. The screw rotation constrains that the OAM aligns along the $z$ direction, i.e. $L_z$. 
Additionally, we note that the orbital texture in the band structure is conceptually different from the net orbital magnetization which is strictly zero at equilibrium because of TRS.

The topological orbital texture originates in the anisotropic hopping of the chiral orbital along the chiral chain. For generality, we consider a right-hand helical chain with the $n$-fold screw rotation, in which the site $i$ has two nearest neighbors, $i+1$ along $+z$ and $i-1$ along $-z$. We set two bases, $p_{\pm}$, at each site, and set the hopping integral as the simple Slater-Koster type\cite{Slater1954}. Among the same orbital ($p_+$ or $p_-$), the hoppings from $i$ to $i \pm 1$ are the same. From $p_+$ to $p_-$, however, the hopping from $i$ to $i+1$ is different from that from $i$ to $i-1$ by a phase factor $e^{-i 4\pi/n}$. If the chirality reverses, the phase $-4\pi/n$ also changes its sign. Take a helix with $n=4$, for example. The inter-orbital hopping is different by a ``--'' sign between the up and down directions. It means that $p_+$ prefers a certain direction in hopping, while $p_-$ prefers the opposite direction. Consequently, the anisotropic inter-orbital hopping induces the orbital texture in the band structure. It also indicates that the orbital texture is as robust as the hopping integral's energy scale, i.e., the bandwidth.

Different from the band structure, the energy level of a finite-size molecule exhibits no dispersion. Nevertheless, we can still use insights from the band structure and regard the molecule wave function as the superposition of the right and left movers from the band structure. Since right and left movers exhibit opposite $L_z$, the molecule displays no orbital polarization at the ground state. When an electron tunnels through the molecule from left to right, it transmits through the right mover channel, as illustrated in \textbf{Figure 1c}. 

We adopt a right-handed helix model to represent a chiral chain in \textbf{Figure 1}. The helix includes three sites in the primitive unit-cell and exhibits the same three-fold screw rotation as the $3_{10}$ helix. We set three $p$-orbitals ($p_{x,y,z}$) on each site and consider the nearest neighboring hopping in a tight-binding model (see more details in the Methods section and SM). As shown in \textbf{Figure 1b}, its band structure also exhibits the same band degeneracy and orbital texture. For a linear chain, in contrast, the orbital polarization is strictly zero at all $k$-points because of the presence of both inversion symmetry and TRS.

(iii) Spin- and orbital-polarized conductance.  We calculate the scattering matrix $S_{nm}$ from the left ($L$) to the right ($R$) lead by the scattering theory and obtain the conductance by the Landauer-B\"uttiker formula~\cite{Buttiker1986}, 
\begin{equation}
    G_{L \rightarrow R} = \frac{e^2}{h}\sum\limits_{n \in R, m \in L} |S_{nm}|^2,
\end{equation}
where $S_{nm}$ is the transmission amplitude from $m$-th eigenstate in the left lead to the $n$-th eigenstate in the right lead (see more details in the SM). In the following discussion, we will use $G$ for $G_{L \rightarrow R}$ if it is not noted specifically.

In both leads, the spin ($S_z=\uparrow \downarrow$) and the orbital ($L_z=\pm, 0$) are conserved quantities because of the disappearance of SOC and the axial rotational symmetry, respectively. Thus, one can classify the conductance into each $S_z$ or $L_z$ channels. Given the non-polarized injection state from the left lead, we estimate the spin and orbital, respectively, polarized conductance of transmitted electrons in the right lead by,
\begin{align}
G_{S_z} &= G_{L \rightarrow R\uparrow}-G_{L \rightarrow R\downarrow} \\
G_{L_z} &= G_{L \rightarrow R+}-G_{L \rightarrow R-},
\end{align}
where $G_{L \rightarrow R S_z(L_z)}$ represents the conductance from the $L$ lead to the $S_z(L_z)$ channel of the $R$ lead. Here, $G_{L \rightarrow R~L_z=0}$ is omitted because the $L_z=0$ state does not contribute to the total polarization. Corresponding spin and orbital polarization ratios are $P_{S_z}=G_{S_z} / G$ and $P_{L_z}=G_{L_z} /G$, respectively. To demonstrate the CISS, we will show that electrons go through the chiral molecule and get spin-polarized, which is caused by the orbital polarization effect, by calculating $G_{S_z}$ and $G_{L_z}$.

In \textbf{Figure 2}, the existence of spin polarization does not require extra dephasing in our calculations. This is because we have three modes in each lead. We find that the spin conductance $G_{S_z}$ is insensitive to the moderate dephasing with strong SOC and turns to decrease when the large dephasing violates the coherence (see \textbf{Figure 2c} and SM). If single-mode leads are employed, we verify $G_{S_z} = 0$ without dephasing (see SM).

We have performed all conductance calculations by our program and verified them with the quantum transport package Kwant\cite{Groth2014kwant}. Related model parameters can be found in the SM.
Our transport codes in Matlab can be download from \href{https://github.com/BinghaiYan/Chirality}{https://github.com/BinghaiYan/Chirality}.

(iv) The dephasing related to the inelastic process was frequently discussed as a necessary condition
~\cite{Guo2012,Gutierrez2012,Matityahu2016,Yang2019,Yang2019b} 
to generate CISS. The single-mode leads employed in these models prohibit the spin current in the presence of TRS \cite{Bardarson2008}.
The existence of multiple modes
in our leads allows the emergence of spin current without introducing the extra dephasing. We note that multiple modes represent the more realistic condition of the transition metal contact, compared to the single-mode model.  However, the existence of the two-terminal MR, i.e. $G_{L \rightarrow R} (+M) \neq G_{L \rightarrow R}(-M)$, requires the dephasing to leak electrons into virtual leads. Otherwise, the charge conservation forces the reciprocity regardless of the mode number in leads. Therefore, {the spin current generation and the MR have distinct relations to the dephasing in CISS}, as discussed in the following sections. In calculations, we introduce the dephasing parameter $i\eta$ as the B\"uttiker virtual probe\cite{Buttiker1986b} equally to each site of the chiral molecule. {The dephasing also includes the lifetime effect of quasiparticles due to scattering by the electron-electron interaction and/or disorders.} As long as the dephasing exists, we find that the nonreciprocal MR manifests the CISS effect in a two-terminal device.
The dephasing term accounts for the dephasing in the linear transport regime and the inelastic relaxation in the nonlinear regime. The dephasing commonly exists due to the energy dissipation and inelastic effects in experiment.

\section*{Supplementary Materials}
\noindent Section S1. Tight-binding model of the helical chain.\\
\noindent Section S2 and Figure S1. Anisotropic hopping along the chiral chain\\
\noindent Section S3 and Figure S2. Model parameters for the transport calculations.\\
\noindent Section S4 Scattering Matrix Method.\\
\noindent Figure S3. Band structures of the lead, the SOC region and the chiral molecule.\\
\noindent Figure S4. Band structures of the chiral molecule with a constant life time.\\
\noindent Figure S5. Band structure and transport calculations for broken $p_{x/y/z}$ degeneracy. \\
\noindent Figure S6. The  total,  orbital  and  spin  conductance  for  chiral  molecules  with  different  length  of  units. \\
\noindent Figure S7. The comparison of the orbital filter effect and orbital polarization effect. \\
\noindent Figure S8. The comparison of the orbital filter effect and orbital polarization effect with different parameters.\\
\noindent Figure S9. Orbital channel-specific conductance of the chiral chain.\\
\noindent Figure S10. Verification of the global Onsager's reciprocal relation.\\
\noindent Figure S11. Influence of dephasing parameter $\eta$ on spin conductance.\\
\noindent Figure S12. Orbital channel-specific conductance of the achiral chain.\\
\noindent Figure S13. The orbital conserved lead and non-conserved lead in the achiral chain device.\\
\noindent Section S4 and Figure S14. The \textit{ab intio} band structure of the chiral materials  \\

\section*{References and Notes}

\noindent \textbf{Acknowledgements:} 
We thank gratefully advice, discussions and help from Ady Stern and Yuval Oreg. We also acknowledge inspiring discussions with Ron Naaman, Yossi Paltiel, Zhong Wang, ChiYung Yam, Lukas Muechler,  Tobias Holder, Raquel Queiroz, Karen Michaelia,  Leeor Kronik, B.J. van Wees, Xu Yang, Shuichi Murakami, Eugene J. Mele and Claudia Felser. B.Y. honors the memory of Shoucheng Zhang who inspired him to investigate chirality in both physics and biology. B.Y. acknowledges the financial support by the Willner Family Leadership Institute for the Weizmann Institute of Science, the Benoziyo Endowment Fund for the Advancement of Science,  Ruth and Herman Albert Scholars Program for New Scientists, the European Research Council (ERC) under the European Union's Horizon 2020 research and innovation programme (Grant No. 815869).\\
\noindent \textbf{Competing Interests:} The authors declare that they have no competing interests.\\
\noindent \textbf{Data and materials availability:}All data needed to evaluate the conclusions in the paper are present in the paper and/or the Supplementary Materials. Additional data related to this paper may be requested from the authors. \\
\noindent \textbf{Author Contributions:} B.Y. conceived and supervised the project. Y.L. and J.X. performed tight-binding and transport calculations. J.K. preformed DFT calculations. B.Y., Y.L. and J.X. wrote the manuscript. All authors discussed and analysed results. \\
\noindent {$^\ast$ These two authors contributed equally to this work.}\\
\noindent {$^\dagger$ Email of correspondence: binghai.yan@weizmann.ac.il}  


\begin{thebibliography}{10}
\expandafter\ifx\csname url\endcsname\relax
  \def\url#1{\texttt{#1}}\fi
\expandafter\ifx\csname urlprefix\endcsname\relax\def\urlprefix{URL }\fi
\providecommand{\bibinfo}[2]{#2}
\providecommand{\eprint}[2][]{\url{#2}}

\bibitem{Siegel1998}
\bibinfo{author}{Siegel, J.~S.}
\newblock \bibinfo{title}{{Homochiral imperative of molecular evolution}}.
\newblock \emph{\bibinfo{journal}{Chirality}} \textbf{\bibinfo{volume}{10}},
  \bibinfo{pages}{24--27} (\bibinfo{year}{1998}).

\bibitem{Weyl1929}
\bibinfo{author}{Weyl, H.}
\newblock \bibinfo{title}{Gravitation and the electron}.
\newblock \emph{\bibinfo{journal}{Proceedings of the National Academy of
  Sciences of the United States of America}} \textbf{\bibinfo{volume}{15}},
  \bibinfo{pages}{323} (\bibinfo{year}{1929}).

\bibitem{Gohler2011}
\bibinfo{author}{Gohler, B.} \emph{et~al.}
\newblock \bibinfo{title}{{Spin Selectivity in Electron Transmission Through
  Self-Assembled Monolayers of Double-Stranded DNA}}.
\newblock \emph{\bibinfo{journal}{Science}} \textbf{\bibinfo{volume}{331}},
  \bibinfo{pages}{894 -- 897} (\bibinfo{year}{2011}).

\bibitem{Naaman2012}
\bibinfo{author}{Naaman, R.} \& \bibinfo{author}{Waldeck, D.~H.}
\newblock \bibinfo{title}{{Chiral-Induced Spin Selectivity Effect}}.
\newblock \emph{\bibinfo{journal}{The Journal of Physical Chemistry Letters}}
  \textbf{\bibinfo{volume}{3}}, \bibinfo{pages}{2178--2187}
  (\bibinfo{year}{2012}).

\bibitem{Naaman2019}
\bibinfo{author}{Naaman, R.}, \bibinfo{author}{Paltiel, Y.} \&
  \bibinfo{author}{Waldeck, D.~H.}
\newblock \bibinfo{title}{{Chiral molecules and the electron spin}}.
\newblock \emph{\bibinfo{journal}{Nature Reviews Chemistry}}
  \textbf{\bibinfo{volume}{3}}, \bibinfo{pages}{250--260}
  (\bibinfo{year}{2019}).
\newblock \urlprefix\url{https://www.nature.com/articles/s41570-019-0087-1}.

\bibitem{BanerjeeGhosh2018}
\bibinfo{author}{Banerjee-Ghosh, K.} \emph{et~al.}
\newblock \bibinfo{title}{{Separation of enantiomers by their enantiospecific
  interaction with achiral magnetic substrates}}.
\newblock \emph{\bibinfo{journal}{Science}} \textbf{\bibinfo{volume}{360}},
  \bibinfo{pages}{1331--1334} (\bibinfo{year}{2018}).

\bibitem{Xie2011}
\bibinfo{author}{Xie, Z.} \emph{et~al.}
\newblock \bibinfo{title}{{Spin specific electron conduction through DNA
  oligomers.}}
\newblock \emph{\bibinfo{journal}{Nano letters}} \textbf{\bibinfo{volume}{11}},
  \bibinfo{pages}{4652--5} (\bibinfo{year}{2011}).

\bibitem{Kettner2015}
\bibinfo{author}{Kettner, M.} \emph{et~al.}
\newblock \bibinfo{title}{{Spin Filtering in Electron Transport Through Chiral
  Oligopeptides}}.
\newblock \emph{\bibinfo{journal}{The Journal of Physical Chemistry C}}
  \textbf{\bibinfo{volume}{119}}, \bibinfo{pages}{14542--14547}
  (\bibinfo{year}{2015}).

\bibitem{Ray1999}
\bibinfo{author}{Ray, K.}, \bibinfo{author}{Ananthavel, S.~P.},
  \bibinfo{author}{Waldeck, D.~H.} \& \bibinfo{author}{Naaman, R.}
\newblock \bibinfo{title}{{Asymmetric Scattering of Polarized Electrons by
  Organized Organic Films of Chiral Molecules}}.
\newblock \emph{\bibinfo{journal}{Science}} \textbf{\bibinfo{volume}{283}},
  \bibinfo{pages}{814--816} (\bibinfo{year}{1999}).

\bibitem{Mishra2013}
\bibinfo{author}{Mishra, D.} \emph{et~al.}
\newblock \bibinfo{title}{{Spin-dependent electron transmission through
  bacteriorhodopsin embedded in purple membrane}}.
\newblock \emph{\bibinfo{journal}{Proceedings of the National Academy of
  Sciences}} \textbf{\bibinfo{volume}{110}}, \bibinfo{pages}{14872--14876}
  (\bibinfo{year}{2013}).

\bibitem{Kettner2018}
\bibinfo{author}{Kettner, M.} \emph{et~al.}
\newblock \bibinfo{title}{{Chirality-Dependent Electron Spin Filtering by
  Molecular Monolayers of Helicenes}}.
\newblock \emph{\bibinfo{journal}{The Journal of Physical Chemistry Letters}}
  \textbf{\bibinfo{volume}{9}}, \bibinfo{pages}{2025--2030}
  (\bibinfo{year}{2018}).

\bibitem{Guo2012}
\bibinfo{author}{Guo, A.-M.} \& \bibinfo{author}{Sun, Q.-f.}
\newblock \bibinfo{title}{{Spin-Selective Transport of Electrons in DNA Double
  Helix}}.
\newblock \emph{\bibinfo{journal}{Physical Review Letters}}
  \textbf{\bibinfo{volume}{108}}, \bibinfo{pages}{218102}
  (\bibinfo{year}{2012}).
\newblock \eprint{1201.4888}.

\bibitem{Gutierrez2012}
\bibinfo{author}{Gutierrez, R.}, \bibinfo{author}{D\'{\i}az, E.},
  \bibinfo{author}{Naaman, R.} \& \bibinfo{author}{Cuniberti, G.}
\newblock \bibinfo{title}{Spin-selective transport through helical molecular
  systems}.
\newblock \emph{\bibinfo{journal}{Phys. Rev. B}} \textbf{\bibinfo{volume}{85}},
  \bibinfo{pages}{081404} (\bibinfo{year}{2012}).
\newblock \urlprefix\url{https://link.aps.org/doi/10.1103/PhysRevB.85.081404}.

\bibitem{Matityahu2016}
\bibinfo{author}{Matityahu, S.}, \bibinfo{author}{Utsumi, Y.},
  \bibinfo{author}{Aharony, A.}, \bibinfo{author}{Entin-Wohlman, O.} \&
  \bibinfo{author}{Balseiro, C.~A.}
\newblock \bibinfo{title}{{Spin-dependent transport through a chiral molecule
  in the presence of spin-orbit interaction and nonunitary effects}}.
\newblock \emph{\bibinfo{journal}{Physical Review B}}
  \textbf{\bibinfo{volume}{93}}, \bibinfo{pages}{075407}
  (\bibinfo{year}{2016}).

\bibitem{Michaeli2019}
\bibinfo{author}{Michaeli, K.} \& \bibinfo{author}{Naaman, R.}
\newblock \bibinfo{title}{{Origin of Spin-Dependent Tunneling Through Chiral
  Molecules}}.
\newblock \emph{\bibinfo{journal}{The Journal of Physical Chemistry C}}
  \textbf{\bibinfo{volume}{123}}, \bibinfo{pages}{17043--17048}
  (\bibinfo{year}{2019}).

\bibitem{Dalum2019}
\bibinfo{author}{Dalum, S.} \& \bibinfo{author}{Hedegård, P.}
\newblock \bibinfo{title}{{Theory of Chiral Induced Spin Selectivity}}.
\newblock \emph{\bibinfo{journal}{Nano Letters}} \textbf{\bibinfo{volume}{19}},
  \bibinfo{pages}{5253--5259} (\bibinfo{year}{2019}).

\bibitem{Yang2019}
\bibinfo{author}{Yang, X.}, \bibinfo{author}{Wal, C. H. v.~d.} \&
  \bibinfo{author}{Wees, B. J.~v.}
\newblock \bibinfo{title}{{Spin-dependent electron transmission model for
  chiral molecules in mesoscopic devices}}.
\newblock \emph{\bibinfo{journal}{Physical Review B}}
  \textbf{\bibinfo{volume}{99}}, \bibinfo{pages}{024418}
  (\bibinfo{year}{2019}).
\newblock \eprint{1810.02662}.

\bibitem{Geyer2020}
\bibinfo{author}{Geyer, M.}, \bibinfo{author}{Gutierrez, R.} \&
  \bibinfo{author}{Cuniberti, G.}
\newblock \bibinfo{title}{{Effective Hamiltonian model for helically
  constrained quantum systems within adiabatic perturbation theory: Application
  to the chirality-induced spin selectivity (CISS) effect}}.
\newblock \emph{\bibinfo{journal}{The Journal of Chemical Physics}}
  \textbf{\bibinfo{volume}{152}}, \bibinfo{pages}{214105}
  (\bibinfo{year}{2020}).
\newblock \eprint{2002.08052}.

\bibitem{Zoellner2020}
\bibinfo{author}{Zoellner, M.~S.}, \bibinfo{author}{Varela, S.},
  \bibinfo{author}{Medina, E.}, \bibinfo{author}{Mujica, V.} \&
  \bibinfo{author}{Herrmann, C.}
\newblock \bibinfo{title}{{Insight into the Origin of Chiral-Induced Spin
  Selectivity from a Symmetry Analysis of Electronic Transmission}}.
\newblock \emph{\bibinfo{journal}{Journal of Chemical Theory and Computation}}
  \textbf{\bibinfo{volume}{16}}, \bibinfo{pages}{2914--2929}
  (\bibinfo{year}{2020}).

\bibitem{Gersten2013}
\bibinfo{author}{Gersten, J.}, \bibinfo{author}{Kaasbjerg, K.} \&
  \bibinfo{author}{Nitzan, A.}
\newblock \bibinfo{title}{{Induced spin filtering in electron transmission
  through chiral molecular layers adsorbed on metals with strong spin-orbit
  coupling}}.
\newblock \emph{\bibinfo{journal}{The Journal of Chemical Physics}}
  \textbf{\bibinfo{volume}{139}}, \bibinfo{pages}{114111}
  (\bibinfo{year}{2013}).
\newblock \eprint{1306.4904}.

\bibitem{Yan2017}
\bibinfo{author}{Yan, B.} \& \bibinfo{author}{Felser, C.}
\newblock \bibinfo{title}{{Topological Materials: Weyl Semimetals}}.
\newblock \emph{\bibinfo{journal}{Annu. Rev. Cond. Mat. Phys.}}
  \textbf{\bibinfo{volume}{8}}, \bibinfo{pages}{337 -- 354}
  (\bibinfo{year}{2017}).
\newblock
  \urlprefix\url{http://www.annualreviews.org/doi/abs/10.1146/annurev-conmatphys-031016-025458}.

\bibitem{Armitage2017}
\bibinfo{author}{Armitage, N.~P.}, \bibinfo{author}{Mele, E.~J.} \&
  \bibinfo{author}{Vishwanath, A.}
\newblock \bibinfo{title}{{Weyl and Dirac semimetals in three-dimensional
  solids}}.
\newblock \emph{\bibinfo{journal}{Rev. Mod. Phys.}}
  \textbf{\bibinfo{volume}{90}}, \bibinfo{pages}{015001}
  (\bibinfo{year}{2018}).

\bibitem{Tassinari2018}
\bibinfo{author}{Tassinari, F.} \emph{et~al.}
\newblock \bibinfo{title}{{Chirality Dependent Charge Transfer Rate in
  Oligopeptides}}.
\newblock \emph{\bibinfo{journal}{Advanced Materials}}
  \textbf{\bibinfo{volume}{30}}, \bibinfo{pages}{1706423}
  (\bibinfo{year}{2018}).

\bibitem{Parameswaran2013}
\bibinfo{author}{Parameswaran, S.~A.}, \bibinfo{author}{Turner, A.~M.},
  \bibinfo{author}{Arovas, D.~P.} \& \bibinfo{author}{Vishwanath, A.}
\newblock \bibinfo{title}{{Topological order and absence of band insulators at
  integer filling in non-symmorphic crystals}}.
\newblock \emph{\bibinfo{journal}{Nature Physics}}
  \textbf{\bibinfo{volume}{9}}, \bibinfo{pages}{299 -- 303}
  (\bibinfo{year}{2013}).

\bibitem{Buttiker1986}
\bibinfo{author}{B\"uttiker, M.}
\newblock \bibinfo{title}{Four-terminal phase-coherent conductance}.
\newblock \emph{\bibinfo{journal}{Phys. Rev. Lett.}}
  \textbf{\bibinfo{volume}{57}}, \bibinfo{pages}{1761--1764}
  (\bibinfo{year}{1986}).
\newblock \urlprefix\url{https://link.aps.org/doi/10.1103/PhysRevLett.57.1761}.

\bibitem{Buttiker1986b}
\bibinfo{author}{Büttiker, M.}
\newblock \bibinfo{title}{{Role of quantum coherence in series resistors}}.
\newblock \emph{\bibinfo{journal}{Physical Review B}}
  \textbf{\bibinfo{volume}{33}}, \bibinfo{pages}{3020--3026}
  (\bibinfo{year}{1986}).

\bibitem{Zwang2016}
\bibinfo{author}{Zwang, T.~J.}, \bibinfo{author}{Hürlimann, S.},
  \bibinfo{author}{Hill, M.~G.} \& \bibinfo{author}{Barton, J.~K.}
\newblock \bibinfo{title}{{Helix-Dependent Spin Filtering through the DNA
  Duplex}}.
\newblock \emph{\bibinfo{journal}{Journal of the American Chemical Society}}
  \textbf{\bibinfo{volume}{138}}, \bibinfo{pages}{15551--15554}
  (\bibinfo{year}{2016}).

\bibitem{Liu2020}
\bibinfo{author}{Liu, T.} \emph{et~al.}
\newblock \bibinfo{title}{Linear and nonlinear two-terminal spin-valve effect
  from chirality-induced spin selectivity}.
\newblock \emph{\bibinfo{journal}{ACS Nano}} \textbf{\bibinfo{volume}{14}},
  \bibinfo{pages}{15983--15991} (\bibinfo{year}{2020}).

\bibitem{Metzger2020}
\bibinfo{author}{Metzger, T.~S.} \emph{et~al.}
\newblock \bibinfo{title}{{The Electron Spin as a Chiral Reagent}}.
\newblock \emph{\bibinfo{journal}{Angewandte Chemie International Edition}}
  \textbf{\bibinfo{volume}{59}}, \bibinfo{pages}{1653--1658}
  (\bibinfo{year}{2020}).

\bibitem{Santra2019}
\bibinfo{author}{Santra, K.}, \bibinfo{author}{Zhang, Q.},
  \bibinfo{author}{Tassinari, F.} \& \bibinfo{author}{Naaman, R.}
\newblock \bibinfo{title}{{Electric-Field-Enhanced Adsorption of Chiral
  Molecules on Ferromagnetic Substrates}}.
\newblock \emph{\bibinfo{journal}{The Journal of Physical Chemistry B}}
  \textbf{\bibinfo{volume}{123}}, \bibinfo{pages}{9443--9448}
  (\bibinfo{year}{2019}).

\bibitem{Ziv2019}
\bibinfo{author}{Ziv, A.} \emph{et~al.}
\newblock \bibinfo{title}{{AFM-Based Spin-Exchange Microscopy Using Chiral
  Molecules}}.
\newblock \emph{\bibinfo{journal}{Advanced Materials}}
  \textbf{\bibinfo{volume}{31}}, \bibinfo{pages}{1904206}
  (\bibinfo{year}{2019}).

\bibitem{Cuniberti2020}
\bibinfo{author}{Dianat, A.} \emph{et~al.}
\newblock \bibinfo{title}{{The Role of Exchange Interactions in the Magnetic
  Response and Inter-Molecular Recognition of Chiral Molecules}}.
\newblock \emph{\bibinfo{journal}{Nano Letters}} \textbf{\bibinfo{volume}{20}},
  \bibinfo{pages}{7077–7086} (\bibinfo{year}{2020}).

\bibitem{Yang2019b}
\bibinfo{author}{Yang, X.}, \bibinfo{author}{van~der Wal, C.~H.} \&
  \bibinfo{author}{van Wees, B.~J.}
\newblock \bibinfo{title}{Detecting chirality in two-terminal electronic
  nanodevices}.
\newblock \emph{\bibinfo{journal}{Nano Letters}} \textbf{\bibinfo{volume}{20}},
  \bibinfo{pages}{6148--6154} (\bibinfo{year}{2020}).

\bibitem{Rikken2001}
\bibinfo{author}{Rikken, G. L. J.~A.}, \bibinfo{author}{FÃ¶lling, J.} \&
  \bibinfo{author}{Wyder, P.}
\newblock \bibinfo{title}{{Electrical Magnetochiral Anisotropy}}.
\newblock \emph{\bibinfo{journal}{Physical Review Letters}}
  \textbf{\bibinfo{volume}{87}}, \bibinfo{pages}{236602}
  (\bibinfo{year}{2001}).

\bibitem{Tokura2018}
\bibinfo{author}{Tokura, Y.} \& \bibinfo{author}{Nagaosa, N.}
\newblock \bibinfo{title}{{Nonreciprocal responses from non-centrosymmetric
  quantum materials}}.
\newblock \emph{\bibinfo{journal}{Nature Communications}}
  \textbf{\bibinfo{volume}{9}}, \bibinfo{pages}{3740} (\bibinfo{year}{2018}).

\bibitem{Zelezny2017}
\bibinfo{author}{Zelezny, J.}, \bibinfo{author}{Zhang, Y.},
  \bibinfo{author}{Felser, C.} \& \bibinfo{author}{Yan, B.}
\newblock \bibinfo{title}{{Spin-Polarized Current in Noncollinear
  Antiferromagnets}}.
\newblock \emph{\bibinfo{journal}{Physical review letters}}
  \textbf{\bibinfo{volume}{119}}, \bibinfo{pages}{187204}
  (\bibinfo{year}{2017}).

\bibitem{Bliokh2007}
\bibinfo{author}{Bliokh, K.~Y.}, \bibinfo{author}{Bliokh, Y.~P.},
  \bibinfo{author}{Savelaev, S.} \& \bibinfo{author}{Nori, F.}
\newblock \bibinfo{title}{{Semiclassical Dynamics of Electron Wave Packet
  States with Phase Vortices}}.
\newblock \emph{\bibinfo{journal}{Physical Review Letters}}
  \textbf{\bibinfo{volume}{99}}, \bibinfo{pages}{190404}
  (\bibinfo{year}{2007}).

\bibitem{Lloyd2017}
\bibinfo{author}{Lloyd, S.~M.}, \bibinfo{author}{Babiker, M.},
  \bibinfo{author}{Thirunavukkarasu, G.} \& \bibinfo{author}{Yuan, J.}
\newblock \bibinfo{title}{Electron vortices: Beams with orbital angular
  momentum}.
\newblock \emph{\bibinfo{journal}{Rev. Mod. Phys.}}
  \textbf{\bibinfo{volume}{89}}, \bibinfo{pages}{035004}
  (\bibinfo{year}{2017}).
\newblock
  \urlprefix\url{https://link.aps.org/doi/10.1103/RevModPhys.89.035004}.

\bibitem{Uchida2010}
\bibinfo{author}{Uchida, M.} \& \bibinfo{author}{Tonomura, A.}
\newblock \bibinfo{title}{{Generation of electron beams carrying orbital
  angular momentum}}.
\newblock \emph{\bibinfo{journal}{Nature}} \textbf{\bibinfo{volume}{464}},
  \bibinfo{pages}{737--739} (\bibinfo{year}{2010}).

\bibitem{perdew1996generalized}
\bibinfo{author}{Perdew, J.~P.}, \bibinfo{author}{Burke, K.} \&
  \bibinfo{author}{Ernzerhof, M.}
\newblock \bibinfo{title}{{Generalized gradient approximation made simple}}.
\newblock \emph{\bibinfo{journal}{Physical Review Letters}}
  \textbf{\bibinfo{volume}{77}}, \bibinfo{pages}{3865} (\bibinfo{year}{1996}).

\bibitem{kresse1996efficient}
\bibinfo{author}{Kresse, G.} \& \bibinfo{author}{Furthm{\"u}ller, J.}
\newblock \bibinfo{title}{{Efficient iterative schemes for ab initio
  total-energy calculations using a plane-wave basis set}}.
\newblock \emph{\bibinfo{journal}{Physical Review B}}
  \textbf{\bibinfo{volume}{54}}, \bibinfo{pages}{11169} (\bibinfo{year}{1996}).

\bibitem{Slater1954}
\bibinfo{author}{Slater, J.~C.} \& \bibinfo{author}{Koster, G.~F.}
\newblock \bibinfo{title}{Simplified lcao method for the periodic potential
  problem}.
\newblock \emph{\bibinfo{journal}{Phys. Rev.}} \textbf{\bibinfo{volume}{94}},
  \bibinfo{pages}{1498--1524} (\bibinfo{year}{1954}).
\newblock \urlprefix\url{https://link.aps.org/doi/10.1103/PhysRev.94.1498}.

\bibitem{Groth2014kwant}
\bibinfo{author}{Groth, C.~W.}, \bibinfo{author}{Wimmer, M.},
  \bibinfo{author}{Akhmerov, A.~R.} \& \bibinfo{author}{Waintal, X.}
\newblock \bibinfo{title}{Kwant: a software package for quantum transport}.
\newblock \emph{\bibinfo{journal}{New Journal of Physics}}
  \textbf{\bibinfo{volume}{16}}, \bibinfo{pages}{063065}
  (\bibinfo{year}{2014}).

\bibitem{Bardarson2008}
\bibinfo{author}{Bardarson, J.~H.}
\newblock \bibinfo{title}{{A proof of the Kramers degeneracy of transmission
  eigenvalues from antisymmetry of the scattering matrix}}.
\newblock \emph{\bibinfo{journal}{Journal of Physics A: Mathematical and
  Theoretical}} \textbf{\bibinfo{volume}{41}}, \bibinfo{pages}{405203}
  (\bibinfo{year}{2008}).

\end{thebibliography}


\end{document}